\documentclass[11pt]{article}
\newcommand{\be}{\begin{equation}}
\newcommand{\ee}{\end{equation}}
\newif\ifpdf
\ifx\pdfoutput\undefined
\pdffalse 
\else
\pdfoutput=1 
\pdftrue
\fi

\ifpdf
   \usepackage[pdftex,urlcolor=blue,colorlinks=true]{hyperref}
   \usepackage[pdftex]{graphicx}
\else
   \usepackage[ps2pdf,pdftitle={PDF},urlcolor=blue,colorlinks=true]{hyperref}
   \usepackage{graphicx}
\fi

\begin{document}
\ifpdf
\DeclareGraphicsExtensions{.pdf, .jpg, .tif}
\else
\DeclareGraphicsExtensions{.eps}
\fi
\title{
Optical feedback and the coupling problem in semiconductor
microdisk lasers}

\author{Jens U.~N{\"o}ckel$^*$\\
Nanovation Technologies, 
1801 Maple Avenue, Evanston, IL 60201}

\date{$^*${\footnotesize
    \href{http://darkwing.uoregon.edu/~noeckel}{\em Current
      address}:\\
Department of Physics,
    University of Oregon, 1371 E 13th Avenue, Eugene, OR 97403\\
{\em Published in} physica status solidi (a){\bf 188}, 921 (2001)}}

\maketitle

\begin{abstract}
  The smaller the size of a light-emitting microcavity, the more important 
  it becomes to understand the effects of the cavity boundary on the optical 
  mode profile. Conventional methods of laser physics, such as the paraxial 
  approximation, become inapplicable in many of the more exotic cavity designs 
  to be discussed here. Cavities in the shape of microdisks, pillars
  and rings can yield low lasing thresholds in a wide variety of
  gain media: 
  quantum wells, wires and even dots, as well as quantum cascade superlattices 
  and GaN. An overview of the experimental and theoretical status is 
  provided, with special emphasis on the light extraction problem.
\end{abstract}

Light emission from microcavities is a problem of great fundamental 
and applied interest. A wide range of possible active 
media can be used to form microcavities, and consequently the 
properties of the microscopic dipoles which generate the light can 
vary significantly. One of our goals is to identify common, 
material-independent features of microcavity emitters, which are 
strongly determined by the geometric dimensions and shape of the 
cavity. Properties of interest are the spectrum of optical modes, their 
internal intensity distribution and the external field profile of the 
emitter. These characteristics in turn determine technological figures 
of merit, such as external quantum efficiency in the case of light 
emitting diodes (LEDs), or pump threshold and maximum output power 
in the case of lasers. 

A problem in the design of LEDs is their 
poor external quantum efficiency, owing to the fact that light
generated within the diode is not easily extracted. This is because total 
internal reflection at the interface between the 
semiconductor ($n_{in}$) and the surrounding 
lower-index medium ($n_{out}$) allows only 
those light rays to escape whose angle of incidence $\chi$ with respect to 
the surface normal satisfies 
\be\label{eq:totalinternal}
\sin\chi<n_{out}/n_{in}\equiv 1/n.
\ee
A first step toward better output
coupling in LEDs is to reduce the vertical dimension (denoted by $z$) 
until a planar cavity is obtained \cite{deneve}. The formation of 
Fabry-Perot modes in the $z$ direction leads to 
redistribution of spectral weight into peaks of the density of 
states $\rho(k)$ \cite{brorson,rkccampillo}. 
As a consequence, the spontaneous emission 
of the microscopic dipoles into the remaining cavity modes is enhanced
according to Fermi's golden rule \cite{slusher,bjoerk,laerinoeckelbook}. 

This can lead to directional LED emission \cite{benisty} because the
resonant cavity modes are spatially anisotropic. In a microcavity {\em
laser}, such as a VCSEL, the desired directional emission has to be balanced
against the additional goal of obtaining laser oscillation at low 
threshold power. These two requirements are incompatible if we 
consider the conventional threshold condition
that gain and loss must balance out: reducing the linear cavity
dimension also reduces the available gain medium, and this eventually 
leads to an explosive increase in pump threshold if the mirror 
reflectivities are kept fixed \cite{chenyariv}. On the other hand,
increased mirror reflectivity means that the available output power is
small. 

Thus, it is desirable both for
LED and laser applications to have as much freedom as possible in the
``mirror design'' of a microcavity, to create the analogue
of the well-known stable, unstable or confocal configurations of
macroscopic laser physics \cite{siegman}. Clearly, vertical layer
structures do not lend themselves to such a level of electromagnetic
engineering. However, the {\em lateral} shape of a semiconductor
microcavity can be chosen arbitrarily. 

To calculate the cavity mode structure, recall that the laser
oscillation condition implies that the wave equation has a solution
with outgoing, but no incoming waves in the far field. This radiation 
boundary condition can be satisfied only at a discrete set of real 
numbers $(k,\,\gamma)$, where $k$ is the wavenumber and $\gamma$ the
exponential gain constant \cite{yariv} of the active medium. This 
eigenvalue
problem can equivalently be stated as a search for complex wavenumbers 
${\tilde k}=k-i\gamma$ satisfying the above boundary condition with
the index $n$ of the cavity at its transparency value;
the solutions are known in scattering theory 
as the {\em quasibound states} \cite{mcbook}. This is 
what we mean by the term ``modes'' of the leaky cavity.

\begin{figure}[t]
\centerline{
\includegraphics[width=10cm]{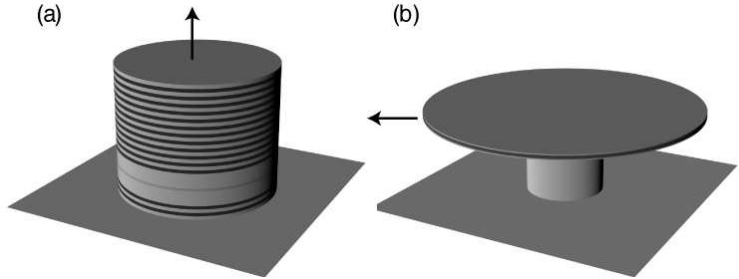}}
\caption{Schematic comparison of VCSEL gemoetry (a) with microdisk 
(b). Arrows indicate predominantly vertical emission in (a) and edge 
emission in (b). Multilayers defining the Bragg mirrors in (a) are 
partially buried. \label{fig:sketches}
}
\end{figure}
A possible lateral geometry for a VCSEL is that of a circular
cylinder, cf. Fig.\ \ref{fig:sketches}. 
It allows comparatively simple modeling because each of its
layers can be considered as a short section of a cylindrical waveguide
for which the modes are known analytically \cite{snyderlove}. In 
particular, there are two types of modes in a cylinder, {\em guided
modes} whose propagation constant $\beta_z$ in the z direction is large, and
{\em leaky modes} with intensity localized close to the cylinder 
side wall. 
When piecing together the fields in the individual layers of the VCSEL
stack \cite{kuszelewicz}, it is not surprising
that we encouter both type of modes in pixelized VCSELs
\cite{ahn}. However, in a finite-sized, three-dimensionally
confined dielectric cavity {\em all} modes are leaky, and
in fact the large-$\beta_z$ modes can turn out to have lower Q than
the ones with small $\beta_z$, when leakage through the planar
mirrors outweighs the lateral radiation losses. This was confirmed in
the threshold behavior of  a circular AlGaAs multiquantum well (MQW) 
cavity \cite{ahn} of vertical thickness $L_z\approx\lambda$ and 
radius $R=15\,\mu$m. 

At large $\beta_z$, the
sidewalls allow waveguiding \cite{djaloshinski} and hence 
there is predominantly
vertical emission, with the planar mirrors controlling the
cavity Q factor. On the other hand, {\em small} $\beta_z$ leads to 
total internal reflection at the planar mirrors 
according to Eq.\ (\ref{eq:totalinternal}), because the angle of
incidence satisfies
$\tan\chi=\sqrt{n^2k^2-\beta_z^2}/\beta_z$,
where $k$ is the free-space wavenumber, and the numerator describes 
the in-plane component of the internal wave vector. The lateral 
leakage is {\em also} controlled by total internal reflection,
provided the light circulates close to the side wall close to grazing
incidence; this is just the ray-optics equivalent of the ring-shaped
intensity patterns which are commonly known as {\em whispering-gallery} 
(WG) modes, due to an acoustic analogue described 
first by Lord Rayleigh \cite{slusher}. 

These ring-shaped modes thus avoid
refractive escape according to Eq.\ (\ref{eq:totalinternal}) at 
{\em all} surfaces of the optical cavity. Hereafter, we 
define $\chi$ to be the angle of incidence with respect to the 
{\em side wall} normal to focus on the lateral emission properties. 
Neglecting absorption in the cavity, the WG mode Q-factor is limited
only because internal reflection at the side walls is in fact 
frustrated by the finite {\em surface curvature}. 
This wavelength-dependent correction to the ray picture can be calculated
straightforwardly within the WKB approximation
\cite{snyderlove,mkchinwkb}, and leads to infinite Q in the
$\lambda\to 0$ limit. The WKB method is here applied to the
radial part of the wave equation governing the cavity field (after 
introducing cylinder coordinates); 
although it is a short-wavelength approximation, it has
been found to provide quantitative results even in microcavities of
dimensions comparable to the wavelength 
\cite{mcbook,slushermohidden,izo}. 

In contrast to a VCSEL mode, WG modes of a circular cavity 
emit predominantly 
sideways, and without any prefered azimuthal (in-plane) direction. 
For such edge emitting devices, Bragg mirrors on top and 
bottom of the cavity may be omitted \cite{weidongzhou}, 
provided the index-difference
between the guiding layer and surrounding cladding is large
enough. The strongest index contrast clearly results if we manage to 
suspend the
guiding layer in air. In this way, we get from the circular VCSEL to 
the {\em microdisk laser} \cite{slusher}: like a thumbtack, the 
semiconductor disk is supported in the center by a pedestal which is
thin enough to have only small or no interaction with the WG
modes. Both optically and electrically pumped microdisks have been
realized \cite{frateschi}. For better heat-sinking at 
room-temperature, one can also form the WG disk on pillars of other
low-index material, e.g. Al$_x$O$_y$ \cite{dssong}.

Rotational symmetry alone is not sufficient to permit exact analytic
solution of the microdisk problem, however. The field of a given 
cavity mode depends on the cylindrical polar angle 
$\phi$ only through $A\,\exp(i\,m\phi)+B\,\exp(-i\,m\phi)$, where 
the integer $m$ is the analogue of a quantum mechanical angular 
momentum index. However, the radial ($r$) and $z$ dependence of the
field remain coupled by the combination of boundary conditions at the
vertical and lateral interfaces \cite{wang}. This makes it necessary
either to perform numerical calculations in the $r-z$ plane
\cite{binjingli}, or to neglect this coupling by making a 
semivectorial approximation \cite{frateschi} which distinguishes
between TE and TM modes. It is in fact confirmed numerically that the
electric field of a given mode is either perpendicular to (TM) or in
the disk plane (TE), up to small corrections. 

If the disk thickness $L_z$ is small enough to satisfy 
$\sqrt{n^2-1} kL_z<\pi$,
then only the fundamental slab-waveguide standing waves in the vertical
direction are supported for both, TE and TM \cite{wang}. The
corresponding range of thicknesses for
semiconductor disks ($n>3$) is therefore approximately
$L_z<\lambda/5$. The TE modes have higher Q
than TM modes because the electric field is better confined in the
vertical direction for TE; reported values for TE modes 
reach as high as $Q=12000$ in a $2\,\mu$m diameter disk with InAs
self-assembled quantum dots (``boxes'') as emitters
\cite{gayral}. More typical Q factors are in the range from
$Q=100\ldots 1000$; circular disks with such quality have been used to
obtain lasing not only from conventional MQW structures
\cite{frateschi} but also from InAs quantum wires in InP
\cite{seassal} where optical pump powers of $\approx 2$mW were found
at room temperature. In GaN, the introduction of a microdisk cavity 
led to a pump threshold reduction by an order of magnitude
\cite{seongsikchang}. When only a {\em single} quantum dot acts as 
the emitter in the disk, lasing is not achieved but highly controlled
emission of {\em single photons} under time-periodiv pumping has been 
demonstrated; the spontaneous-emission enhancement in the WG
microcavity here serves to reduce the timing jitter \cite{imamoglu}.

The fact that a microdisk laser 
emits from its edge could be thought to cause extreme vertical
spreading of the output radiation, if we recall that a slab waveguide
of thickness $d$ has a divergence angle of order $\sim \pi d/\lambda$:
with disks as thin as $0.2\,\mu$m emitting at $\lambda=1.55\,\mu$m,
the spreading should be prohibitive. However, this is not
the case  \cite{binjingli,tdlee}. The reason is that the far-field of
a microdisk laser {\em cannot} be obtained from the near-field on its
surface by a simple Fourier transformation, as would be the case for
the planar output mirror of a VCSEL \cite{forchel}. Rotational
symmetry implies conservation of the $\phi$-``angular momentum'' $m$ 
between the interior and exterior fields, and this impresses a
centrifugal potential onto the free-space propagation of the emitted
light in the $r-z$ plane \cite{mcbook}. This in turn favors radial over
vertical propagation, reducing the weight of large $\beta_z$ in the
emission below that expected for a Gaussian beam. A rough estimate 
yields \cite{binjingli} a spreading angle 
(FWHM) $\Delta\theta\approx 2/\sqrt{m}$. 

The largest $m$ admitted by a microdisk cavity of radius $R$ and
effective slab index $n$ can be estimated from
the condition Eq.\ (\ref{eq:totalinternal}). One can show that 
semiclassically, a mode with azimuthal index corresponds to rays with
angle of incidence satisfying
$m=n\,kR\,\sin\chi$,
and since the maximum of $\chi$ is $\pi/2$, we find that high-Q modes
must have $m\le n\,kR$. As a consequence, $\Delta\theta$ becomes
smaller for larger disk radii, provided they lase on WG modes
close to grazing incidence at the side walls. 

Whereas vertical focusing of the emission around the disk plane is
thus surprisingly efficient, there is no easy prescription for inducing
focused {\em azimutal} emission within this plane. As was recognized 
early on in the pioneering experiments of Slusher {\em et al.}, the 
rotational symmetry of the disk has to be destroyed \cite{levislusher}. 
In the infinite
space of possible shape deformations, it is challenging to find simple
design rules that create preferential emission while at the
same time preserving the desired high Q-factor. One type of shape
perturbation that can never be ruled out is side wall roughness. It
turns out, however, that both wavelength and Q factor of WG modes are
quite insensitive to small random perturbations of the circular 
geometry \cite{binjingli2}. Additional perturbative effects can be
caused by the pedestal \cite{baba}. 

\begin{figure}[t]
\centerline{
\includegraphics[width=10cm]{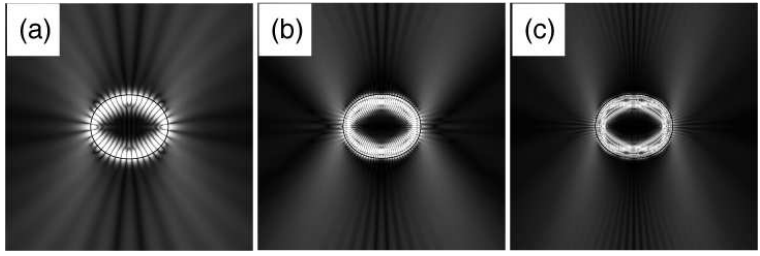}}
\caption{Three chaotic WG modes with similar intensity profiles
(plotted in gray scale) for the same shape deformation and effective 
refractive index $n=2$. We measure the complex wavenumbers of these
states in relation to the mean cavity radius, $R$: by definition, the
area of the oval is $A=\pi\,R^2$. The real and imaginary parts of 
${\tilde k}$ are then 
(a) $kR=11.97$, $\gamma R=2.45\times 10^{-3}$, 
(b) $kR=27.28$, $\gamma R=1.80\times 10^{-3}$, and 
(c) $kR=44.19$, $\gamma R=2.84\times 10^{-3}$. These are the same
states described in Ref.\ \cite{nature}; they are chosen
such that they semiclassically correspond to approximately the same family of
chaotic ray trajectories.\label{fig:modes}
}
\end{figure}

When the geometry of the disk is distorted to such an extent that
perturbative treatments \cite{leung} break down, 
numerical modeling becomes much more complex because $m$ is no longer 
a ``good quantum number'' by which the modes can be labeled. 
Motivated by the success of the WKB method mentioned above, it was suggested
early on that short-wavelength approximations can provide valuable
insights into the mode structure and emission patterns of microlasers
\cite{mcbook,optlett94,mekis,optlett96}. The main observation is that
the internal ray dynamics of a non-circular disk is in general
{\em chaotic}, so that methods from the fields of classical and quantum
chaos theory become applicable. Three chaotic WG modes are shown in 
Fig.\ \ref{fig:modes}, for a wide range 
of different wavelengths. At the deformation
considered here, chaotic ray motion is the dominant mechanism for
escape from the cavity, as opposed to diffractive
($\lambda$-dependent) effects. This is evidenced by the fact
that $\gamma$ in Fig.\ \ref{fig:modes} depends only weakly on wavenumber
$k$. Likewise, one can
identify wavelength-{\em independent} features in the internal and
external intensity profiles; the emission directionality at large
deformations is one of these
characteristics, which to lowest order can be explained without
recourse to the wave nature of light. 

The first experimental tests of these
predictions were performed on lasing dye microdroplets whose shape is
generically oval \cite{mekis}. When rotational symmetry is absent, 
$\chi$ in Eq.\ (\ref{eq:totalinternal}) is not conserved between 
reflections, so that diffusion of photons away 
from the WG region tends to occur; the modal decay rate then 
generally increases with deformation because even WG rays can reach 
Eq.\ (\ref{eq:totalinternal}) after sufficiently many loops in the
cavity. 

However, the Q-spoiling due to large deformation is {\em less severe}
than expected from ray considerations alone, as was pointed out in 
Ref.\ \cite{nature}. An interference effect called ``dynamical
localization'' actually suppresses decay rates in a manner analogous to
Anderson localization in disordered solids \cite{dynloc}, thus making high-Q 
asymmetric cavities more feasible: the advantages of
asymmetric shapes for directional emission can then outweigh the
loss in cavity Q.

Experimental realizations of non-circular microdisks \cite{backes} 
confirm another important
prediction of the chaos analysis \cite{optlett94}: 
oval shapes favor WG lasing only as long as the boundary remains 
everywhere {\em convex}, i.e., has no points of vanishing curvature.
The lateral emission from oval disks is found to emanate from 
the regions of highest curvature in semiconductor disks. The light
furthermore escapes {\em tangentially} to the surface, as expected for rays
at the critical angle of incidence $\chi_c$ just satisfying 
Eq.\ (\ref{eq:totalinternal}).

A striking exception to this rule is expected in similar structures
made of polymers \cite{dodabalpur} 
or other low-index materials: the chaotic 
ray dynamics in such oval cavities \cite{sschang} can 
counter-intuitively {\em prevent} WG rays from reaching $\chi_c$
at the high-curvature points. This ``dynamical eclipsing'' occurs because 
ray chaos is not random, but deterministic; its underlying structure 
can in particular rule out certain $\chi$ for WG trajectories 
approaching the high-curvature points. By changing the material 
(refractive index) of the cavity to bring $\chi_c$ into this 
``forbidden'' range, the emission pattern for the same shape 
as in Fig.\ \ref{fig:modes} then changes significantly. 

In contrast to smooth oval deformations, one can also introduce 
notches or projections in the side walls, but this tends to increase
vertical spreading \cite{backes2}, thus reducing the expected focusing 
of the laser. Recalling the dependence of vertical
spreading $\Delta\theta$ on 
angular momentum, we should aim for smoothly deformed microdisks 
emitting {\em near the critical angles of
incidence}, yet as anisotropic as possible. At large deformations, 
cavity modes with this property can have internal intensity 
distributions that are qualitatively different from WG modes. 

This can be illustrated by a 
series of experiments on oval lasers, incrementally
ramping up the deviation from circular shape. Such measurements have
been performed in quantum-cascade microcylinders in the
shape of a flattened quadrupole \cite{gmachl}. The devices were
intended for high power and hence had larger area than the examples
above: keeping the minor axis at $b=50\,\mu$m diameter and increasing 
the major axis to $a=80\,\mu$m, it was found that beyond a threshold
deformation $a\approx 70\,\mu$m, an exponential increase in the 
maximum achievable output power set in. The lasing mode was identified
as belonging to a closed ``bowtie'' ray pattern which does not exist
below the threshold deformation. 

The effect of chaos in this experiment is to spoil all cavity modes
overlapping the gain region, {\em except} for the bowtie modes. This
reduces the density $\rho$ of available modes, while at the same time the
internal intensity profile of the remaining modes becomes more focused
in the electrically pumped center of the structure. The four vertices
of the bowtie pattern meet the side walls near critical
incidence, resulting in directional emission. The combined 
power of the four emission lobes of up to 10 mW was $\approx 1000$ 
times larger than that of the circular device in the test series. 

A special feature of the quantum-cascade system that makes it amenable
to this near-critical mode is that (unlike the previous examples) it
strongly favors emission in TM polarization. This is important because
in a high-index cavity the critical angle for total internal
reflection is close to the {\em Brewster angle}; therefore, 
near-critical TE modes suffer from much higher lateral emission than 
their TM counterparts (the reverse holds for the vertical losses). 

A reduced $\rho$ permits reduced lasing thresholds, and the
lesson learned from the bowtie laser is that chaos can reduce the mode
density even in relatively large-area cavities. This is because 
the average of $\rho$ is not determined by the cavity volume, but by the 
available {\em phase space volume} for long-lived ray
trajectories. The phase space occupied by a ray trajectory is determined
by the real-space area it explores, combined with the range of $\chi$
it spans. We have recently shown \cite{noeckelstone,laeri} that in a WG-cavity 
of mean radius $R$ (cf. Fig.\ \ref{fig:modes}) and index $n$, the average number of 
long-lived (non-overlapping) modes per unit wavenumber interval is
\be\label{eq:wgdensity}
\langle
\rho(k)\rangle\,dk=
(1/4)n^2\,k\,R^2\,\left[1-(2/\pi)\,\left(
\arcsin(1/n)+(1/n)\sqrt{1-(1/n)^2}\right)\right]\,
dk.
\ee
The first term is determined by the cavity area, but the corrections
for small $n$ show how the openness of the cavity reduces the density
of modes by expanding the classical escape window 
Eq.\ (\ref{eq:totalinternal}) which depends on $\chi$. 

In conclusion, phase space reduction and hence lower threshold can be 
achieved by shrinking the real area of
the cavity, e.g. with holes in the disk \cite{backes2} or ring
structures \cite{ho}, {\em or alternatively} by
engineering the side wall deformation to obtain  stable modes such as 
the bowtie while simultaneously destablilizing most others. Concepts
from nonlinear dynamics provide design guidelines for the latter
approach. The rich structure provided by the ray dynamics in
chaotic cavities makes it possible to design shapes for laterally 
directional emission, which makes it possible to collect 
the emitted light more efficiently than in rotationally symmetric 
structures.

\end{document}